# Myths about new ultrahard phases: Why materials that are significantly superior to diamond in elastic moduli and hardness are impossible?

Vadim V. Brazhkin [1,*] and Vladimir L. Solozhenko [2]

[1] *Institute for High Pressure Physics, Russian Academy of Sciences, 142190 Troitsk, Russia*

[2] *LSPM–CNRS, Université Paris Nord, 93430 Villetaneuse, France*

Reports published in the last 25 years on the synthesis of carbon-based materials significantly superior to diamond in hardness and elastic properties have been critically examined, and three groups of recently appearing myths have been analyzed. The first group concerns the possibility of producing materials with bulk moduli much higher than that of diamond. The second group regards to «experimentally measured» hardness significantly higher than that of diamond. Myths of the third group state that quantum confinement effects supposedly provide «theoretical» foundations for a several-fold increase in the hardness of covalent materials. The fundamental impossibility of synthesizing materials with elastic moduli noticeably exceeding those of diamond under normal conditions has been demonstrated. The problems of hardness measuring have been discussed; it was shown that the formation of obstacles for motion of dislocations can allow increasing the measured hardness of superhard materials by 20–40%. It was demonstrated that other hypothetical ways for hardness increase, e.g. owing to quantum confinement have no any real physical grounds. The superior mechanical properties of diamond are due to reliably established physical laws. Accordingly, any statements on the possibility of obtaining materials with elastic characteristics and/or hardness several times higher than the corresponding values for diamond cannot be considered as scientifically reliable.

## I. Introduction

Before the beginning of the 1990s, works on the synthesis and study of new superhard materials constituted a quite respectable field of Materials Science. The relation between elastic properties of materials and their atomic and electron density, degree of anisotropy of the latter, and the structure features (the coordination number, in particular) have been established to that time (see [1] and references therein). It was also recognized that such mechanical properties as hardness, fracture toughness and abrasion resistance can be considered only as qualitative comparative characteristics that depend on the measurement method and conditions. In particular, the results of indentation hardness tests depend on the load, loading time, shape of the indenter, method and quality of processing of the material surface under study, degree of elastic recovery of an imprinting, etc. (see [1, 2] and references therein). No one doubted that diamond and its isoelectronic analog, cubic boron nitride, have the highest elastic moduli among all materials. We recall that diamond has bulk modulus of 445 GPa, shear modulus of 530 GPa, and hardness of 80–120 GPa [1, 3]. New superhard materials, i.e. materials with elastic moduli and hardness exceeding the corresponding values for $Al_2O_3$ sapphire or hard alloy (sintered tungsten carbide with binder) were sought since the 1950s among compounds of light elements (Be, B, C, N, and O), as well as among carbides, borides, oxides, and nitrides of heavy metals (Ta, Mo, Nb, W, Re, Os, Cr, etc.) [1]. It was also completely understood that industrially important mechanical characteristics of any material is significantly dependent on its structure and defect state at the nanolevel (Hall–Petch effect) [1, 2]. However, in contrast to metallic alloys, for which the nanocrystalline state can be relatively easily reached, nanocrystalline superhard materials were unattainable until the end of 1990s, except of TiN–NbN and TiN–VN multilayer nanostructures [2, 4], for which hardness increased by a factor of 1.5–2 compared to the bulk materials was observed.

---

[*] Corresponding author: brazhkin@hppi.troitsk.ru

In the last 25 years a number of publications appeared claiming synthesis of new ultrahard materials (primarily based on carbon) with hardness and bulk moduli exceeding the corresponding values for diamond by a factor of 2–4 [5–7]. These materials were strongly disordered carbon composites obtained by high-temperature treatment of $C_{60}$ fullerenes at pressures of 13–14 GPa. In some cases, the reported values of bulk moduli and hardness of those materials reached 1700 GPa [5] and 300 GPa [7–9], respectively! Those works were widely criticized [2, 3, 10–12] and did not receive further development. In 2014 the formation of similar superhard fullerite-based materials at moderate (6.5 GPa) pressures and room temperature under high shear stresses using carbon sulfide $CS_2$ as a catalyst of three-dimensional polymerization of fullerite was reported [13].

In 2004, an open letter to the Materials Science community has been published [14], where we recalled that diamond remains and will remain an unprecedented material in terms of shear modulus under normal conditions and called upon all researchers to be careful when reporting the results on «hardness above the diamond» value. After that, almost no new publications claiming the production of materials with elastic moduli and hardness several times exceeding those of diamond appeared during the next 5–7 years.

At the same time, a significant advance occurred in the synthesis and study of nanocrystalline diamond (and later nanocrystalline cubic boron nitride) [15–22]. The nanocrystalline state of diamond was obtained using various initial materials (graphite, fullerites, nanotubes, and soot) [17, 19], and sizes of crystal grains in the ultrahard nanocrystalline bulks varied from 5 to 100 nm [15–22]. It appeared that nanocrystalline diamond and cubic boron nitride have higher hardness, as well as much higher wear resistance and thermal stability than the corresponding single crystals. It was found that the highest mechanical characteristics of nanocrystalline diamond can be reached at the grain size of 10–20 nm [16–18]. The measured Knoop hardness of such nanocrystalline bulks makes 130–140 GPa, which is 10–40% higher than the hardness of single-crystal diamond.

The advance also occurred in theoretical calculations of mechanical properties of superhard materials. Modern computers allow high-accuracy (to 1–2%) *ab initio* calculations of elastic characteristics of superhard materials and their ideal strength under various deformations, even beyond the linear elastic region but with a significantly worse (up to tens percent) accuracy (see [23–27] and references

4therein). Since the hardness of materials is an ill-defined parameter, calculations are usually performed for «ideal» hardness defined as the theoretical shear strength corresponding to certain load on an indenter [28]. With a further load increase, the local lattice stability loss, as well as the formation of dislocation loops or twin intergrain boundaries, is observed even in the ideal case [28]. Empirical formulas for hardness calculations based on correlations between mechanical properties and elastic moduli, bond covalence, optical gap width, etc. were also improved [29–32] (and references therein). Although such formulas are not exact, they are convenient for express estimates of mechanical properties of hypothetical phases, and are very actively (but often uncritically) used up to date.

New reports on the synthesis of materials with elastic moduli and hardness several times higher than those of diamond actively appeared in the last 5–10 years. Moreover, some researchers attempted to propose a certain «theoretical» foundation for new experimental results. The aim of this paper is to analyze scientific adequacy of such claims.

## II. Myth I: Materials with giant elastic moduli exceeding that of diamond

This concept is the simplest for denouncement. Elastic moduli are determined by the interparticle interaction and can be calculated *ab initio* with a high accuracy. It is obvious that elastic moduli are higher in a material with a higher binding energy, higher atomic and electron densities and, correspondingly, smaller interatomic distances.

### 1. Bulk modulus

The bulk modulus is associated with the binding energy density in a material. For a degenerate electron gas, the bulk modulus ($B$) is unambiguously related to the electron density $\rho$ as $B \sim \rho^{5/3}$. In real materials, the total energy $E$ is a function of the volume: $E = E_0 f(V/V_0)$, where $E_0$ and $V_0$ are the energy and volume at ambient pressure, respectively. Then, the bulk modulus at ambient pressure has the form [33]

$$B_0 = E_0/V_0 f''(V/V_0)_{V=V_0} = C\, E_0/V_0. \tag{1}$$

In a number of model cases, e.g. for a potential where the cohesion energy can be represented in the form $E = A/V^n - B/V^m$, $f''(V/V_0)_{V=V_0} = mn$, i.e. for the Lennard-Jones potential, $f'' = 8$. This is in good agreement with experimental data for condensed rare gases and a number of molecular materials, where indeed $B_0 \approx 8\, E_0/V_0$ [1].

The bulk modulus in metals, covalent materials, and partly in ionic compounds will be determined mainly by the density of valence electrons. It was found long ago that for many ionic compounds $B \sim 1/V \sim \rho$, where $V$ is the specific volume [34], and for covalent materials with diamond-like structure $B \sim 1/d^{3.5} \sim \rho^{1.17}$, where $d$ is the interatomic distance [35]. The interatomic interaction is not certainly reduced to the simple interparticle potential, but the coefficient $C$ in Eq. (1) varies only slightly for all materials with a «strong» interparticle interaction (covalent materials, metals, and ionic compounds) [1]. In particular, in the carbon subgroup, the coefficient *C* for superhard covalent diamond differs only by 10% from that for soft metallic lead [36]. A semiconducting grey tin and metallic white tin have the same value of C [36]. Thus, the so-called «covalence» (or the inhomogeneity of the electron density) hardly contributes to the bulk modulus. The compressibility of a material is determined primarily by the spatially averaged electron density [1–3]. It can be concluded that the highest bulk modulus of diamond is due to its highest electron (and atomic) density among all materials. In turn, highest density of diamond is associated with its position in the periodic table is in the middle of the first filled row in view of a small ionic radius and four valence electrons [37]. The density of valence electrons indirectly affects the total energy. As a result, all materials with the «strong» interparticle interaction on average exhibit the correlation $B_0 \sim \rho^\alpha$, where $\alpha \approx 1.25$ [2], which is close to the relation $B \sim \rho^{1.17}$ previously established for covalent materials [35].

The above consideration concerns materials with small anisotropy. The strong anisotropy of the compressibility along different directions will be observed for quasi-one- or quasi-two-dimensional materials. The bulk modulus will certainly be lower than it would be for an isotropic solid with the same electron density. For example, the bulk modulus of three-dimensionally bound isotropic $sp^2$ carbon phases is 200–300 GPa, which is an order of magnitude (!) higher than the $B \approx 20$–30 GPa value for anisotropic graphite, although the linear compressibility of graphite in the basal plane is lower than that for diamond (see [1–3] and references therein). The problem of topological rigidity is close to the problem of anisotropy [38, 39]. At a decrease in the average coordination number (*Z*) down to *Z* = 2.4–2.7 (depending on the bonds topology), the covalent structure loses rigidity, and the elastic moduli (bulk modulus, in particular) decrease sharply, and the structure decomposes into quasi-one- and quasi-two-dimensional regions [3].

Thus, diamond has the highest atomic density and the highest density of valence electrons among all known materials and these parameters are responsible for its highest bulk modulus. Is it possible to obtain a material more incompressible than diamond? This cannot be absolutely excluded. Moreover, the electron density of some transition metals such as W, Re, Os, and Ir, as well as their compounds, is close to the «diamond» value, and the bulk moduli of these phases range from 300 to 430 GPa (for diamond, *B* = 445 GPa). The existence of stable or metastable compounds with a lower compressibility is not excluded. However, in any case the bulk moduli of such materials at ambient pressure cannot exceed the bulk modulus of diamond by more than 10–20% [1–3]. It should be noted that carbon has some allotropes several percent denser than diamond, and some of these phases according to calculations can be kinetically stable at ambient pressure [1, 40]. The bulk modulus of such predicted allotropes can also be several percent higher than the «diamond» value [40], however, materials with bulk moduli much higher than that of diamond are highly unlikely.

Experimental studies of carbon materials synthesized from fullerites at high pressure, as well as announcements of super high moduli and/or hardness of such materials, were stimulated and provoked mostly by a series of curious theoretical works [41–43]. According to the estimates made, the bulk modulus of a single $C_{60}$ molecule is formally twice as high as the diamond value. Then, it was erroneously assumed that a close packed crystal can be synthesized from touching $C_{60}$ molecules. The formal bulk modulus of such a crystal is 620–720 GPa, which is higher than the diamond value by a factor of about 1.5. These estimates are trivial because the formal density of a single fullerene molecule is 6.3 g/cm$^3$, which is almost twice as high as diamond density, and the density of a hypothetical crystal consisting of touching $C_{60}$ molecules is also higher than the diamond value by a factor of 1.5 [12]. However, such a crystal cannot exist in principle because carbon atoms from neighboring $C_{60}$ molecules should be spaced by an impossibly small distance (smaller than 0.5 Å); i.e. molecules cannot touch each other [12]. The density and bulk modulus of any polymerized structure based on fullerene molecules are noticeably lower than the diamond values. It is noteworthy that smaller carbon molecules or clusters will formally have higher densities and bulk moduli. In particular, the density and bulk modulus of a single hypothetical carbon cube $C_8$ and a single hypothetical carbon tetrahedron $C_4$ are higher than the respective diamond values by factors of 9 and 40 (!), respectively [12]. However, these parameters are not referred to a bulk carbon material; the notion of elastic moduli is inapplicable to individual atoms or molecules.

We now discuss experimental and theoretical evidences of synthesis of materials significantly superior to diamond in bulk moduli. The first such report appeared in 1998 [5]. For bulk amorphous and nanocrystalline carbon materials obtained from $C_{60}$ fullerite at high pressures and high temperatures, the velocities of longitudinal and transverse waves were measured by acoustic microscopy. The bulk and shear moduli were calculated from these data. Anomalously high bulk moduli up to 1700 GPa, which is four times (!) higher than the diamond value, were reported for some samples. At the same time, the densities of these carbon phases are 10–20% lower than the density of diamond. Moreover, the measured shear modulus for these samples is lower than that of diamond by a factor of 2–3 (!), which leads to nonphysical Poisson's ratio ($v > 0.4$), that is characteristic of e.g. rubber or soft metals like lead rather than of covalent carbon material. In addition, such shear moduli would mean that the hardness of these carbon phases should be much more lower than the diamond value. However, these materials were reported to be harder than diamond [7–9] (see below). As shown in [2, 44], the reasons for possible errors in the bulk modulus determination from the data on sound velocities are the anisotropy (nanotexturing) and inhomogeneity of the samples properties. Such an anisotropy appears very often and is due to the temperature and pressure gradients during the synthesis. Nevertheless, unrealistic value of bulk modulus of 1700 GPa was mentioned in the subsequent works, both experimental [7, 45] and theoretical [46, 47].

Some reports on super high bulk moduli were based on X-ray diffraction data under pressure. An error of 150–250% of those data was due to strongly non-hydrostatic conditions in the experiment [2]. In particular, bulk modulus of 600 GPa for the three-dimensionally polymerized fullerite phase was reported in [6], whereas later studies of the similar samples under more hydrostatic conditions gave $B = 280$ GPa [10] and $B = 220$ GPa [11]. The same fabrication conditions (13-14 GPa, 900ºC) have been used in [48] to prepare large samples of the fullerite-based superhard materials. The moduli of these samples (both bulk and shear) have been measured by the Brillouin scattering technique [48].

The reasonable values $B = 368$ GPa and $G = 375$ GPa were obtained with "normal" corresponding value of the Poisson's ratio $v = 0.12$ [48].

Indirect estimates of bulk moduli of amorphous or strongly disordered phases were made (and continue to be made) from the measured pressure dependence of Raman frequencies which results in overestimated bulk modulus values (for instance, $B = 585$ GPa [13]). In the case of a crystal, knowing the Grüneisen parameter for a given mode, one can estimate the compressibility from the pressure dependence of the corresponding frequency. For the strongly disordered state this is fundamentally impossible because compression strongly changes the shape of a broad Raman band, and the notion of a fixed Grüneisen parameter is inappropriate. Furthermore, the amorphous state can have a lower compressibility of individual interatomic bonds than that in the corresponding crystal and, simultaneously, a higher compressibility of the entire amorphous sample [49]. This is due to the permanent distortion and transformation of the structure of amorphous network under compression [49]. Finally, the cited Raman studies were also performed under strongly non-hydrostatic conditions, which could lead to a significant (150–200%) error in the determination of the pressure derivative of Raman frequencies.

The synthesis of carbon samples with higher (by 10%) bulk modulus and higher (by 0.2–0.4%) density than corresponding diamond values was reported only in [19]. These samples were also obtained from $C_{60}$ fullerite at high pressures and high temperatures, and had the structure of aggregated needle-like diamond nanocrystals with the thick amorphous grain boundaries. When determining the X-ray density and bulk modulus (also by X-ray diffraction under pressure), only lines corresponding to well crystallized diamond grains were taken into account. X-ray amorphous regions (about half the sample volume) were disregarded. Highly likely this is a reason for the erroneous statement on the «densest and least compressible form of carbon» [19]. High shear stresses held in the aggregated diamond nanorods; in this case, X-ray data also provide an overestimated bulk modulus value although the measurements were performed under quasi-hydrostatic conditions. Recently the same authors have reported the similar data on bulk moduli of nanocrystalline diamond microballs $B = 489$ GPa [50] and $B = 482$ GPa [51]. The estimated X-ray density of the balls was 3.54 g cm$^{-3}$ (0.5% higher than diamond). The reason of these erroneous measurements is just the same as discussed above.

The picosecond ultrasonic spectroscopy study of similar diamond nanopolycrystals synthesized from graphite in [52] also indicated an increase in the longitudinal sound velocity by 3% as compared to the average value for single-crystal diamond. Diamond has a cubic lattice but splices of twinned nanocrystals can exhibit apparent anisotropy (favorable direction for both crystal growth and twins appearance which is related to temperature and pressure gradients during the synthesis). In this case, ultrasonic data will not correspond to the standard average values for the polycrystal. This is indirectly confirmed by the fact that an increase in the longitudinal sound velocity was not observed for some samples, moreover, the sound velocity in some samples was lower than that in the single crystal [52].

Thus, all evidences of synthesis of carbon materials significantly superior to diamond in bulk modulus are based either on erroneous experimental data or on their improper interpretation.

Curious reports on the synthesis of carbon materials with bulk moduli of 600, 800, and even 1700 GPa [5, 6, 13] could be ignored in a discussion, but the possibility of synthesizing such materials was supposedly explained in a series of recent computer simulations [46, 47, 53]. According to theoretical calculations, the bulk modulus of any carbon phase with lower density than that of diamond should be less than bulk modulus of diamond. For this reason, the authors of [46, 47] assumed that a composite material with elastic moduli and density above the respective diamond values can be obtained from two carbon phases, each with elastic moduli and density below the respective diamond values at ambient pressure. A particular implementation of such composites involves nanograins of three-dimensional polymers of $C_{60}$ fullerite that are strongly compressed inside the diamond matrix. At ambient pressure the initial polymerized phase has density of 3.1 g/cm$^3$ and calculated bulk modulus of 340 GPa. The authors of [46, 47] believe that grains of this phase with density up to 4-5 g/cm$^3$ can be located in a compressed state inside the diamond matrix. We note that such density change corresponds to an external pressure of 200-500 GPa on the grains; correspondingly, the negative pressure (tensile stresses) with the same magnitude should acts on the diamond matrix. The maximum tensile strength of diamond is ~100 GPa. At higher tensile stresses, the plastic flow of diamond occurs [23–25, 54]. Consequently, such a degree of compression of fullerite polymers is unattainable. However, the main fault of these calculations is of different nature. According to the theory of elasticity [55], when a compressed region is formed inside the matrix, the total change in the volume of the matrix and inclusion made of the same materials is exactly zero. Shear stresses appear in the system and lead to the same volume changes with opposite signs in the compressed region and matrix. As a result, the density of a hypothetical composite based on the same material should be equal to the density of the initial unstressed mixture of phases with the same phase ratio. This is invalid in the case of different materials. At the synthesis of a stressed composite material, a more compressible material undergoes a larger change in the density and bulk modulus than a denser and less compressible material. However, materials with different bulk moduli cannot provide the synthesis of a composite with bulk modulus higher than the moduli of both constituents. Roughly speaking, a material with the density and moduli above the respective parameters of steel cannot be fabricated from rubber and steel. It is impossible to create the composite from 2 carbon phases with an average density higher than the density of the denser component. The real maximum pressures that can act on polymerized fullerite in the diamond matrix in the elastic regime are below 100 GPa. In the case of 1:1 fullerite-to-diamond ratio, the fullerite will be compressed by about 15%, and the diamond matrix is expanded by 10%. Consequently, in any case the density and bulk modulus of the composite cannot exceed 3.4 g/cm$^3$ and 400 GPa, respectively. The authors of [46, 47] did not present the average density of the entire modeled system. This density (as well as the bulk modulus of the entire system) should be intermediate between the densities of diamond (3.5 g/cm$^3$) and polymerized fullerite (3.1 g/cm$^3$). If this is not the case (according to [46, 47]), the modeled system was effectively under very high pressure (200-300 GPa) and very high density (about 4.5 g/cm$^3$), and the authors incorrectly used periodically boundary conditions (actually, very recently the authors admitted their errors). It would be useful to consider the limiting case of a stressed material where the matrix and grains are made of the same material, e.g. diamond. Diamond grains inside the diamond matrix can be compressed, the diamond matrix will be expanded, the total change in the volume will be zero, and the bulk modulus of such a diamond will decrease slightly (the negative contribution from expanded regions will dominate over the positive contribution from compressed regions).

An ideologically similar approach was used to explain an increase in the moduli of nanocrystalline diamonds as compared to the single-crystal values [52]. The contribution from stressed twin boundaries was taken into account, and it was shown that high concentration of these twin boundaries can increase elastic moduli, but only by several tenths of percent. The results vary within several percent depending on the interatomic potential used in the calculation. Such a small increase is indeed possible for shear or Young's modulus in certain directions in an anisotropic system, but is forbidden by the thermodynamic laws for the bulk modulus. The bulk modulus in any complex stressed state of the sample at ambient pressure can only decrease. In [50], the possible increase in the bulk modulus of nanocrystalline diamond was also explained by the complex anisotropic shear stress in some grains. However, as mentioned above, shear stresses cannot increase the density and bulk modulus (but can reduce them). It should be noted that recent similar study of the elastic moduli of similar nanocrystalline diamonds [56] did not reveal any enhancement of moduli of nanocrystalline substances in comparison with single-crystal diamond.

Thus, any theoretical foundations for the possibility of carbon materials with bulk moduli significantly higher than the diamond value but with lower density are absent.

## 2. Shear and Young's moduli of ultrahard materials and anisotropic systems

Shear ($G$) and Young's ($E$) moduli are more important for operational characteristics of materials such as the strength, hardness, fracture toughness and wear resistance because shear stresses are present in any mechanical test. Several (up to 21) elastic constants exist for a single crystal depending on symmetry. Here we discuss primarily the shear and Young's moduli of polycrystalline materials. These moduli are also related to the bond energy density and, consequently, to the atomic and electron density of the material [1, 2]. However, in contrast to the bulk modulus, shear and Young's moduli are determined not only by the average density of valence electrons, but also by the degree of its spatial localization, which cannot affect the determination of the bulk modulus, as mentioned above. The spatial distribution of the electron density is obviously important because any shear strain changes this distribution, and the so-called angular rigidity appears. The dependences of the shear modulus on the average electron density for different materials lie between $G \sim \rho$ and $G \sim \rho^{5/3}$ [2]. A high degree of covalence and, correspondingly, a tangent stiffness can increase the shear modulus by a factor of 1.5–2 as compared to the more uniform distribution of electron density, e.g. in metals [1, 2]. In particular, covalent compounds SiC and $B_4C$ have higher shear moduli than W and Re metals although their bulk moduli are lower [1, 2]. Metallic white tin is superior to semiconductor grey tin by 20% both in density and bulk modulus, but its polycrystalline shear modulus is lower by a factor of almost 1.5 [1, 2]. Covalent phases with a nonuniform distribution of electron density are usually formed for structures with small coordination number. However, as in the case of bulk modulus, the topological stiffness of a system is a critical factor [38, 39]. As a result, covalent structures with high shear moduli have coordination numbers in the range from 3 to 6, and the coordination number 4 (as in the diamond lattice) is close to optimal (see discussion in [1–3]). The degree of covalence can be formally characterized by Pugh's ratio $k = G/B$ or Poisson's ratio $(3 - 2G/B)/(6 + 2G/B)$. Among metals with high coordination number, only Be has a high Pugh's ratio and, correspondingly, a low Poisson's ratio because of very small ionic radius and strongly inhomogeneous electron density.

Thus, the shear modulus (as well as Young's modulus) is also determined primarily by atomic density of the material and density of valence electrons. However, in contrast to bulk modulus, shear modulus can vary by a factor of 1.5–2 (sometimes up to 3) depending on the degree of spatial inhomogeneity of electron density (degree of «covalence» of bonds).

Diamond has the highest shear modulus (and highest hardness) even among hypothetical carbon materials, including materials with higher density and higher bulk modulus [40]. In contrast to the bulk modulus, reports on the shear moduli above the diamond value are almost absent in both experimental and theoretical works. Theoretical calculations demonstrate the possibility of an increase in elastic moduli by tenths of percent only for nanocrystalline diamond with numerous twin boundaries [49], but these data were quite ambiguous and very sensitive to the used effective empirical interatomic potentials. Moreover, the precise measurements of elastic properties of the polycrystalline diamond by both GHz ultrasonic interferometry and sphere resonance method [56] demonstrated the bulk and shear modili of nanopolycrystalline diamond are identical to those of single-crystal diamond.

Although the average polycrystalline shear and Young moduli of diamond are (and will remain) certainly the highest among all materials (including hypothetical ones), the situation for anisotropic structures is much more interesting and promising. It is known that a graphite single crystal has a higher energy of interatomic bonds along planes and, correspondingly, a lower compressibility than diamond. Chains of carbyne (linearly coupled quasi-one-dimensional carbon allotrope) have the highest uniaxial compression modulus. Fibers based on nanotubes or carbynoid structures can have not only the highest uniaxial moduli but also the correspondingly highest ultimate tensile strength. Not only a several-fold increase in the hardness but also an increase in Young's modulus along certain directions by tens percent are observed in many anisotropic multilayer nanostructures (see [2, 4] and references therein).

We note that all elastic moduli are defined for small strains (linear approximation of the theory of elasticity). In the case of hydrostatic compression, the bulk modulus depends on pressure because of large strains (changes in the volume). In the case of shear strains, the situation is more complex. At the critical strain, the crystal lattice loses stability [23–28]. The theoretical shear strength σ was previously considered within simplified models; in particular, the famous Frenkel formula $\sigma \approx G/2\pi$ was derived almost a century ago [28]. The theoretical shear strength of diamond is estimated as ~85 GPa. Such estimates are widely used to date. At the same time, modern computers allow *ab initio* calculations of the ideal shear strength and uniaxial tensile strength of a crystal. Such calculations are actively performed in the last 20 years. Despite a high accuracy of *ab initio* calculations for predictions of the cohesion energy and electron density distribution, the scatter of values of the theoretical shear strength is still rather large. The main reason for this is the necessity of a very careful inclusion of change in the spatial electron density distribution at large strains (this change can be ignored when calculating elastic moduli in the linear elastic region). In particular, for the ideal shear strength for different directions in a defect-free diamond single crystal, different authors give values from 60 to 200 GPa (see [23–25] and references therein). The most accurate calculations yield values from 95 to 115 GPa [26]. Semi-quantitative experimental measurements give critical shear stresses of ~130 GPa [54]. We note that the ideal strength (stress-induced loss of stability) of crystals for various types of deformations, as well as elastic moduli, is determined only by the structure and interparticle interaction; i.e. it is an unambiguously defined microscopic parameter. This means that such parameters are the same elastic moduli or their combinations, but in a strongly nonlinear region of large strains.

In summary, no reliable experimental data and no theoretical foundations exist for synthesis of materials significantly superior to diamond in elastic moduli at ambient pressure. Problems in experimental measurement of elastic moduli are due to small samples sizes, their disordered and heterogeneous structure, anisotropy and texturing at nano- and microlevels, non-hydrostatic conditions of experiments, etc. Theoretically, the highest elastic moduli of diamond are due to its highest valence electron density (the energy density, in fact) and a high degree of localization of the electron density. The synthesis of any composite material can hardly change the situation.

### III. Myth II: materials much harder than diamond

The hardness of «ultrahard» materials is the most intricate myth complicated for analysis. The main reason for this is the indefiniteness of the notion of «hardness». Hardness is a qualitative rather than quantitative characteristic, and it depends not only on the properties of a material, but also on the measurement method and interpretation of the results.

#### 1. Hardness: definitions and methods for the measurement

According to one of the most commonly accepted definitions, hardness is the degree of resistance of a material to penetration of a harder indenter or scratching by it. In most cases, this definition additionally implies «plastic deformation». Sometimes, hardness is defined as «hardness in the elastic regime»; this parameter is unambiguously determined by elastic moduli of the material (stiffness), the indenter shape, and the load.

The historically first method of measurement (estimation) of hardness was the «comparative scratching» method, sclerometry. In the Mohs scale of mineral hardness corresponding to the ability of minerals to scratch each other, they are characterized by the conditional hardness from 1 (talc) to 10 (diamond). This method became out-of-date to the middle of the past century in view of the appearance of the quantitative methods for hardness measurement by indentation. However, the hardness of microsamples obtained at very high pressures is still estimated in terms of absence or appearance of scratches on flat diamond anvils [8, 57, 58]. We emphasize that scratching in fact means microindentation, and a sample particle can plastically deform the surface of an anvil even if the hardness of the sample is lower than the hardness of the flat anvil by a factor of 3−5 (see discussion below).

The most widely used methods of hardness measurement involve employment of various indenters made of superhard materials. Hardness is measured by Brinell (spherical indenter), Rockwell (conical indenter), Vickers, Berkovich or Knoop (pyramidal indenters with different profiles) methods. Hardness at indentation is defined as ratio of the load to the impression surface area (sometimes, to the area of impression projection, to the impression depth, or to the impression volume). According to the standard definition, hardness corresponds to the average pressure in the region of plastic deformation. «Recovered hardness» is defined when the impression is analyzed after unloading. «Unrecovered hardness» is defined when the impression area or, more often, the penetration depth of the indenter is analyzed directly under loading. Unrecovered hardness is measured, in particular, during nanoindentation. It is clear that correct measurement of recovered hardness requires the presence of a significant plastic strain of the sample. Otherwise, no impression remains after unloading at indentation in the elastic regime; this will formally mean infinite hardness. We emphasize that such absurd results can be obtained not only for hard materials, but also, e.g. for soft rubber.

According to the above discussion, the most correct hardness data can be obtained by means of *in situ* detection of indenter displacement at nanoindentation. The initial stage of nanoindentation always corresponds to the elastic regime, and Young's modulus of material can be estimated with a high accuracy from the data obtained [28]. At a further indenter penetration, and an increase in stresses, a sharp elastic-plastic ("pop-in") transition occurs at a certain depth in a nanocontact due to the nucleation of dislocations and/or twin boundaries in the region under the imprint [28]. If this imprinting nanoregion is free of mobile defects, the onset pressure of the transition corresponds to the «ideal» material hardness. This parameter is unambiguously determined by the material moduli in the strongly nonlinear region of large strains (see above). At the nanoindentation-induced elastic-plastic transition, the homogeneous avalanche-like nucleation and motion of dislocations occur, which usually lead to a sharp decrease in the measured hardness. At the "pop-in" transition in metals, several hundreds or thousands dislocations appear, whereas only several tens dislocations arise in brittle hard materials, e.g. in sapphire [28]. Although *in situ* nanoindentation provide the most reliable information, the use of nanoindenters to study superhard materials is limited for several reasons. First, because of high price and complexity of application, nanoindentation devices cannot be widespread hardness testers for all research groups. Second, nanoindentation imposes stringent requirements on the quality of the sample surface and indenter tip. Finally, it is the most important that the nanoindentation study of materials comparable in hardness with diamond can detect only elastic segments of loading curves, and usually the indenter is destroyed before the elastic-plastic transition in the sample [28].

All these methods for the measurement of indentation hardness were originally developed only for the cases where the hardness and elastic moduli of the indenter are at least several times higher than the respective values for the studied material. It is suggested that change in the indenter size and shape at its elastic deformation under loading can be neglected and, the more so, the plastic deformation of the indenter is inacceptable altogether. This means that the «ideal» indenter should be incompressible and inflexible. In the case of a diamond indenter, these conditions are almost satisfied for most of the studied materials; i.e. the diamond indenter can be considered as «ideal». However, when the elastic and mechanical properties of the indenter and sample under study are comparable, indentation is only a comparative qualitative rather than a reliable quantitative method of measurements. Diamond indenter measurements of the hardness of a material with characteristics close to or slightly exceeding the diamond values can give values several times higher than those obtained with the use of a hypothetical absolutely hard incompressible indenter.

Hardness is measured in three loading ranges: macro, micro, and nano. The nanorange limits only the penetration depth of the indenter, which should be smaller than 0.2 μm. It is noteworthy that a value of 0.1–0.2 μm corresponds to the standard rounding radius of the sharpest triangular Berkovich pyramid used in nanoindentation testers [28]. As a result, the apex of a pyramid at nanoindentation serves as a spherical indenter. The microrange (microhardness) restricts a load to 2 N and the penetration depth of the indenter much larger than 0.2 μm for the pyramidal or conical part of the indenter «to be involved in operation». The characteristic rounding radius of Vickers and Knoop pyramid apex is 0.2–0.4 μm. The macrorange limits the load from 2 N to 30 kN.

The measured hardness depends on various parameters, primarily, on the load on the indenter. For a spherical indenter, the average pressure under the impression increases with the load (Hertz task), and the measured hardness increases correspondingly. For the conical and pyramidal indenters, the pressure under the impression is independent of the load [59], and the measured recovered hardness decreases with increasing load and is almost saturated at high loads. Because of an almost spherical rounding near the apex of a pyramid or a cone, the measured hardness can first increase with the load and, then, start to decrease (for the conical or pyramidal part of the indenter). The reason for a decrease in the measured hardness with increasing load is a decrease in the relative fraction of the elastic strain energy of the material as compared to the energy spent on the formation of a new surface under plastic deformation [1, 2]. The hardness measured for superhard materials is usually saturated starting with large loads of 10–100 N [1]; i.e. standard microindentation testers are inapplicable for their study. We note that the impression area is not strictly proportional to the load even in the plastic regime (see [2] and references therein). Empirical Meyer's law is usually valid: $P \sim d^n$, where $P$ is the load, $d$ is the linear size of the imprint, and $n$ is the exponent usually lying in the range from 1.5 to 2 [60]. The exponent $n$ for superhard materials is noticeably different from 2: $n \approx 1.5$ for diamond and $n \approx 1.4$ for nanocrystalline diamond [16]. Thus, a small decrease in the measured hardness with increasing load is observed even in the plastic region. In particular, a slow decrease in the measured hardness for superhard boron suboxide $B_6O$ is observed up to a load of 200 N [14].

The hardness measured by the same method is determined not only by the load but also by other parameters such as quality of the sample surface under study, the method of preparation of a smooth flat surface (polishing, etching, ion beam processing, etc.), and the properties of the indenter (type of diamond, sharpness of faces and apex, etc.). The indenter loading and unloading rates, loading time, and time interval between unloading and measurement are often important. The process of plastic deformation has a certain kinetics: new dislocation loops, twin boundaries, and other defects can be generated at long-term loading. Many materials demonstrate viscoelastic properties, and the impression left after unloading «is healed» (disappears partially or completely) with time. Studies of materials with moduli and hardness comparable with the respective diamond values face additional complicating circumstances, which will be discussed below. Finally, we note that measurements of hardness are usually performed under normal conditions. At high temperatures, the plasticity (i.e. the simplicity of formation and mobility of dislocations and twin boundaries) of brittle solid materials increases sharply, which can result in a strong decrease in the measured hardness. In particular, the Vickers hardness of the diamond single crystal measured at a temperature of 1400 K decreases down to 15-20 GPa; i.e. at this temperature diamond becomes softer than silicon carbide SiC (~25 GPa) [2].

The above consideration concerns the hardness measurement of a particular sample. Measurements of different samples of the same material will give hardness depending on the grain sizes, defects, and stresses in each particular sample. This problem will be discussed below.

## 2. «Theoretical» models of hardness

Information presented in the preceding section seems to unambiguously indicate that hardness is not a clearly defined quantitative characteristic. Hardness of the material measured by a single method can differ by tens and even hundreds percent. Consequently, no theory of hardness and no theoretical models of hardness are possible altogether. Furthermore, hardness is almost useless as a technological characteristic. It is certainly very desirable that the elements of a tool for material processing (cutting, sharpening, polishing, drilling, etc.) are harder or, at least, not much softer than the processed material. However, this is the only requirement. Other high technological properties of solid materials, such as fracture toughness, wear resistance, thermal stability, and chemical resistance to the processed material are much more important for industrial applications. However, these more important technological characteristics are always in «shadow» of hardness, particularly, in research articles and popular scientific literature. Moreover, interest in new superhard materials only grows, and new empirical formulas for the determination of hardness appeared recently. The main reasons are the transparency of the notion «hardness» (ability to scratch, to indent) and the active use of the words «hard» and «soft» in everyday language (contrary to e.g. the terms «wear resistance» or «fracture toughness»). Furthermore, the creation of the industry of synthetic superhard materials (diamond and cubic boron nitride) is one of the greatest stages of the technological revolution in the 20th century. The current total world production of synthetic diamond and cubic boron nitride amounts thousands of tons per year (two orders of magnitude larger than the mining of natural diamonds). The world industry of superhard materials involves hundreds billion dollars. For this reason, it is attractive to use the terms «new superhard material» or «material harder than diamond» in articles and grant proposals. It is also reasonable that physicists and chemists specialized in Materials Science are tempted to develop a more strict theoretical foundation for indistinct notion of hardness.

The most obvious idea is to find correlation between hardness measured by a certain method and the elastic moduli, which are distinct characteristics of a material. Such correlations were studied for almost a century (see [1–3, 61] and references therein). Correlation between hardness and bulk modulus exists but with numerous exclusions [1, 3]. Many metals have very high bulk modulus and a relatively low hardness. More certain correlations are observed between hardness and shear modulus, as well as between hardness and Young's modulus [2, 3, 62]. This is not accidental because all methods for the hardness measurement are associated with the shear strain of a material [1]. Moreover, as mentioned above, under loading of an ideally sharp undeformable conical or pyramidal indenter (with a tip angle of $2\varphi$), the pressure in the elastic region («ideal» hardness) is determined only by Young's modulus of the material and is independent of the load [1]:

$$P = E \cot \varphi/(2 - 2v^2), \qquad (2)$$

where $E$ is Young's modulus and $v$ is Poisson's ratio. Such an «ideal» hardness in the elastic region for diamond lies in the range from 200 to 250 GPa depending on the tip angle of the used indenter.

All empirical theories of hardness developed to the end of the last century are in fact reduced only to these correlations between the measured hardness and elastic moduli. However, as mentioned above, it was tempted to calculate the hardness of predicted hypothetical materials on a more reliable quantitative foundation. In particular, the semiquantitative consideration of indentation in [30] gives an estimate of $H \approx 0.151G$ for hardness. Such a coefficient of proportionality indeed satisfactorily describes correlation between the measured hardness and shear modulus [2, 3, 30]. This «novel» formula is «not far» from the century-old Frenkel estimate $H \approx G/2\pi$ of the theoretical shear strength. A hardness of 81 GPa calculated for diamond is slightly below the previously accepted values. At the same time, it is known that among different materials with close shear moduli, the material with a higher Pugh's ratio $k = G/B$ has a higher measured hardness. The approximate relation $H \sim k^2 G$ was obtained for hardness (see [30] and references therein). However, this correlation is not used directly in this form. Instead of it, the relation $H = 2(k^2 G)^{0.585} - 3$ was obtained from a great body of experimental data [30]. This relation cannot certainly be treated as «theoretical». It is a purely empirical formula; furthermore, it has an incorrect dimension (all numerical coefficients depend on units in which the quantities $B$ and $G$ are given; the empirical formula was obtained for the case where the quantities $B$, $G$, and $H$ are measured in GPa). We note that «exact» formulas for hardness provide a paradoxical situation: hardness values predicted in most of calculations are presented with accuracy to tenths of a percent, whereas the experimental hardness of a material even within single method cannot be determined with an error smaller than 10%; moreover, the measurement error for very hard materials is significantly larger.

An actively developed field of the «theory» of hardness for covalent materials concerns correlation of the maximum shear strength of the ideal crystal and the semiconductor gap. This empirical correlation was found by Gilman more than 50 years ago [63, 64], and the demerits of this approach are also well known for a long time [65, 66]. However, these empirical estimates are magically transformed in the last ten years to the «strict theory» of hardness [29, 31, 32, 67]. We discuss in detail the disadvantages of these so-called «theories» in the section devoted to the myth of quantum confinement.

It is well known experimentally that the measured hardness (and sometimes elastic moduli) can depend on the size of crystal grains in a polycrystal, as well as on the concentration of various defects (dislocations, point defects, and chemical impurities) [1–3, 68]. Real crystals usually have a noticeable concentration of dislocations or twin boundaries, which ensure plastic deformation already at moderate shear stresses. At the same time, grain boundaries and other defects and impurities at their high concentration hinder the motion of dislocations. As a result, the mechanical characteristics (hardness, strength) for most materials increase with a decrease in grain size (Hall–Petch effect, see [1] and references therein). In a certain grain size range, hardness is described well by the empirical formula $H = H_0 + C \cdot d^{-1/2}$, where $d$ is the grain size (Hall–Petch law). Beginning with grain size of about 5–10 nm, a further decrease in the grain size results in a decrease in the mechanical characteristics (inverse Hall–Petch effect). This is due to active sliding and rotation of very small grains with respect to each other, as well as to the appearance of coherent twin boundaries [69]. The Hall–Petch effect was taken into account in «theoretical» expressions for hardness in a number of recent works [32], where it was assumed that the total hardness is the simple sum of the «theoretical» hardness for the single crystal and various contributions from grain boundaries (Hall–Petch effect) and other defects. Such an approach is partially justified only if empirical expressions fitted to experimental data are used for the hardness of single crystal (as in [30]). However, the hardness of the ideal defect-free single crystal should be considered as the theoretical shear strength, at which the crystal lattice loses stability (in fact, this corresponds to the nucleation of the first dislocation loop and the beginning of plastic deformation). In this case, no additional contributions from grain boundaries and other defects to hardness should be added: the Hall–Petch effect only approaches hardness to the ideal value.

## 3. Experimental studies of materials «harder» than diamond

In this section, we analyze the reported experimental studies of materials much harder than diamond single crystals.

As mentioned above, the first wave of such reports appeared in the end of 1990s and was initiated primarily by Blank's group [7–9]. This group synthesized superhard carbon phases from $C_{60}$ fullerite via the polymerization and partial destruction of molecules at high pressures and high temperatures. In first experiments, it was observed that phases synthesized from fullerite at very high pressures leave scratches on diamond anvils in a high-pressure cell. Ability to scratch diamond anvils is often referred up to date to as a «proof» of an extraordinarily high hardness of newly synthesized materials [57, 58]. This conclusion is obviously erroneous: under high loading, the sharp edges of powder grains scratch the flat surface of another material even if hardness of the former material is lower than that of the latter material by a factor of 3–5 (see discussion below). More recently, superhard materials from fullerite were obtained in the form of bulks with a size of several millimeters, which allowed quantitative studies by conventional methods. The synthesized carbon phases had strongly disordered (quasi-amorphous) structure and density of 3.1–3.35 g/cm$^3$. It is noteworthy that the authors of [7–9] correctly mentioned that indentation measurements of hardness are hardly reliable in the case of comparable mechanical characteristics of the sample and indenter, and proposed their own method of quantitative sclerometry [8, 9]. This method is based on a comparative study of scratches on various materials by means of an atomic force microscope. It was found that scratches on the surface of a diamond single crystal under the same experimental conditions are deeper and have a larger number of microcracks than those on the surface of «superhard fullerites» [8,9]. It means that hardness or (and) fracture toughness of the "superhard fullerites" is higher than those for diamond. Semiquantitative sclerometry measurements of new samples gave hardness values from 170 to 300 GPa [7–9]. However, we note that similar measurements of the hardness of various faces of diamond gave values from 140 to 170 GPa, which is almost twice as high as the commonly accepted values (80-110 GPa). The authors of [7–9] recognized that the absolute values of the estimated hardness should be treated with caution. It could only be stated that new materials, «superhard fullerites», are comparable with diamond in hardness, and anvils made of the new material leave impressions on the faces of the hardest diamond single crystals (type IIa). The authors of [8, 9], understanding ambiguity of hardness measurements focused on the study of other, more important technological characteristics such as wear resistance. It was established that samples of new carbon materials significantly exceed industrial polycrystalline diamond in wear resistance and fracture toughness [9].

In 2003 synthesis of nanocrystalline diamond was reported by Irifune et al. [15]. The nanocrystalline bulks (several cubic millimeters) were obtained by the direct solid-phase transformation of graphite to diamond at high (above 15 GPa) pressures and high (above 2300 K) temperatures. In the next 15 years, this group systematically studied the structures and properties of nanocrystalline diamonds synthesized from various initial carbon materials (graphite, soot, fullerites, nanotubes, etc.) with wide variation of the synthesis pressure and temperature, and the samples size was increased to a centimeter scale [16-18]. Later a detailed study of the new nanocrystalline diamonds have been performed, see e.g. [70-74].

Diamond bulks with crystal grain sizes from 5 to 100 nm (depending on the initial carbon material and synthesis parameters) and mainly lamellar (plate- or needle-like) grain morphology were synthesized [16–18, 70-74]. The hardness of these materials was studied by the «classical» indentation methods. It was established that Vickers and Berkovich indenters are destroyed at the first loading even under moderate loads, which indicates a very high hardness of diamond nanopolycrystals [16]. Reliable reproducible impressions were obtained only with Knoop indenter and only at loads below 7 N. These impressions were studied in detail by atomic force and electron transmission microscopes [16–18, 70-72]. The majority of the measurements were performed only at a single load of 4.9 N. At this load, the load dependence of hardness does not yet approach a plateau; consequently, the measured hardness is significantly conditional. At the same time, this method allows quantitative comparison of the corresponding parameters for different superhard materials. It was established that the maximum hardness is observed for the diamond polycrystals with a plate-like structure and a grain size of 10–20 nm obtained from graphite [16–18, 70-72]. The Knoop hardness (at a load of 4.9 N) of such materials lies in the range of 120–145 GPa [16–18, 70-74]. The hardness of the hardest diamond single crystals (type IIa) measured by the same method was no more than 130 GPa, and the hardness of ordinary diamond single crystals (type Ib) was 105 GPa and below [16]. The hardness of diamond nanopolycrystals obtained from other carbon materials (including $C_{60}$ fullerites) measured by the same method with Knoop indenter ranged from 80 to 120 GPa depending on the grain size and morphology [16, 17]. The nanoindentation study of ultrahard materials up to a load of 0.25 N (penetration depth up to 450 nm) reveals an almost completely elastic character of loading [16]. We recall that the elastic-plastic ("pop-in") transition was not observed in previous nanoindentation studies of ordinary diamond crystals (penetration depth is usually below 100–200 nm) [28]. Although diamond nanopolycrystals at room temperature are only insignificantly superior in hardness to type IIa diamond single crystals, this difference becomes significant at high temperatures. In particular, the hardness of diamond nanopolycrystals at temperatures above 1000 K is almost twice as high as the hardness of single crystals [18]. This is obviously due to an easier plastic deformation and a higher mobility of dislocations in single crystals at high temperatures. Diamond nanopolycrystals also have significantly higher wear resistance and fracture toughness than both single crystals and industrial polycrystals [16–18, 70-74].

The next announcement of the synthesis of the material noticeably harder than diamond was made by a group from Geophysical Lab, where large diamond single crystals were grown by the chemical vapor deposition (CVD) method [75]. These single crystals (after annealing at a temperature of 2200 K and a pressure of 7 GPa) had Vickers hardness above 160 GPa, which was determined at high loads of 10–30 N [75]. Low-quality impressions of the indenter were studied on an optical microscope. The authors stated that the indenter was not destroyed and left impressions even at high loads, but it could be used only once or twice. The Knoop hardness of these samples was not measured. The real hardness of these single crystals is apparently lower than the hardness of diamond nanopolycrystals synthesized by Irifune et al. because the indentation of the nanopolycrystals at loads above 5 N always resulted in the destruction of Vickers indenters [16]. It can be assumed that the real hardness of the annealed CVD diamonds was below 130 GPa.

The mechanical characteristics of the aggregated diamond nanorods synthesized by direct transition of $C_{60}$ fullerite at very high pressures by German group [19] were also studied in detail [20]. It appeared that Vickers hardness of these samples cannot be measured reliably, and the Knoop hardness at a 5 N load varied from 90 to 115 GPa, in agreement with the results by Irifune et al. [16–18, 70-74]. The difference in the microstructure of the samples was possibly due to the synthesis conditions (different temperature and pressure gradients). Similar to the nanopolycrystals obtained by Irifune et al., the aggregate of diamond nanoneedles had the anomalously high wear resistance and fracture toughness [20].

In the last five years, a number of reports on the synthesis of ultrahard materials based on diamond and cubic boron nitride (cBN) were published by Chinese groups [76–81]. Diamond polycrystals with a nanotwinned structure (nanotwinned diamond) were synthesized at 20 GPa and 2200 K from onion carbon nanoparticles [77]. Further, diamonds with nanotwins were obtained from other initial materials (nanotubes, fullerite, etc.) [77–81].

Those works were started by the production of nanotwinned cBN with extraordinary mechanical properties [76] simply following the previously reported high-pressure synthesis of ultrahard nanocrystalline cubic boron nitride [22]. We recall that the shear modulus of cubic boron nitride is below (by 30%) only the diamond value [1]. At the same time, the measured hardness of cBN single crystals is 2–2.5 times lower than that of diamond, apparently because of the easier formation of dislocations and their easier propagation. By forming a nanostructure in cubic boron nitride, its Vickers hardness was previously increased up to 80–85 GPa [21, 22]. The authors of [76] claimed that cubic boron nitride with nanotwins has a higher Vickers hardness of 108 GPa and Knoop hardness of 78 GPa. These results were reasonably criticized [82]. The authors of [76] have tried to answer this criticism [83] but their reply was not very convincing.

The authors of [77] claimed that the nanotwinned diamond samples with average twins thickness of 5 nm had Vickers hardness of 175–203 GPa (at 5-N load) and Knoop hardness of 168–196 GPa. The hardness was measured at loads up to 8 N, and the fracture toughness was measured at loads of 9.8 and 19.6 N. The Vickers indenter was not destroyed even at such high loads [77] (in contrast to tests of diamond nanopolycrystals by Irifune et al. [16]). The highest hardness of nanotwinned diamonds was explained by the quantum confinement effects [77] (as previously it was made for nanotwinned cubic boron nitride [76]). It was predicted that a further reduction of the twins sizes can increase hardness up to 400 GPa [77]. However, the authors did not present any information on the roughness of the surface of studied samples and on the rounding radius of the indenter tip, as well as images of impressions. It is noteworthy that the penetration depth of Vickers pyramid at the stated hardness and 5-N load was 0.8–1 μm, which is comparable both to roughness of the polished surface (0.3–1 μm) and to rounding radius of the indenter (0.2–0.4 μm). In review [78], the same authors also announced the Vickers hardness much higher than 200 GPa, but the images of impressions were also not presented and discussed. The impressions (Vickers pyramid, load of 9.8 N) obtained on the surface of superhard samples studied by means of optical and atomic force microscopes were presented in more recent work [79]. It was reported that Vickers hardness of some samples reaches 200–250 GPa. The impression depth after indentation at a load of 9.8 N was 0.2 μm, which indicates a huge degree of elastic recovery. We note that the penetration depth of Berkovich indenter at *in-situ* nanoindentation of diamond nanopolycrystals was 0.45 μm at a load of only 0.25 N [16]. The quality and profile of impressions presented in [79] are such that hardness can be estimated only with an accuracy of no better than 30–50%. A noticeable plastic strain in the sample is really observed only near the apex and edges of the Vickers pyramid. Indentation at these loads is not yet obviously saturated (significant plastic strain is absent).

Finally, in the subsequent studies of nanotwinned diamonds [80, 81], it was claimed that both Vickers and Knoop hardnesses reached 300–400 GPa, and the possibility of unprecedented hardness of 600 GPa (!) for the smallest twins was predicted.

Reports of Chinese groups on the synthesis of nanotwinned diamonds with hardness to 300-400 GPa and expected hardness up to 600 GPa were met with justified criticism and caused a lot of questions. First, the possibility of indentation of such ultrahard materials by an ordinary diamond indenter is highly questionable. To explain their results, the authors of [78, 79, 83] stated that the sample under indentation is subjected to tensile shear stresses, whereas the pyramidal indenter is subjected to compressive shear stresses. For this reason, the authors believe that the indenter can leave an impression on a material four or five times harder than the indenter itself. This is hardly valid: the profiles of both shear and compressive stresses in indenter and sample are very complex and vary in the process of indentation. At the initial stage of indentation, the spherical part of the indenter operates, the contact is flattened, and, then, the pyramidal part begins to penetrate into the sample. In addition, it is necessary to take into account friction between indenter and sample, which is also different for different segments of the indenter. The indenter produces the highest stress at a small depth under the contact spot and near the sharp edges of the pyramid. It is known experimentally that the plastic flow of diamond indenter (as well as of diamond anvils with a close apex angle) occurs at pressures of 180–190 GPa [54], which is twice as high as the pressure of the onset of plastic deformation of flat diamond surface under the indenter; i.e., the indenter in the elastic regime can leave an impression on the sample with hardness twice as high as that of the indenter. If the plastic deformation and fracture of indenters are not observed, this means that contact pressures (and the hardness of the sample) were much below 180 GPa.

The pressure under indenter can obviously continue to increase with load after the onset of plastic deformation, and the maximum pressures can additionally be increased by a factor of 2–2.5 [54], but the plastic strain (or fracture) of the indenter in this case should be very high and have to be observed. Thus, the flat surface can be scratched by a microsample whose hardness is lower by a factor of 3–5. Consequently, scratches on the surfaces of diamond anvils cannot indicate an anomalously high hardness of the sample leaving traces (see above).

The quality of impressions in [79] is hardly satisfactory, but the authors used the standard definition of hardness in terms of the diagonal of the «recovered» impression (after unloading). It is obvious that the degree of elastic recovery is anomalously high in all studies of these nanodiamond materials, and the size of the recovered and unrecovered impressions can be very strongly different. Furthermore, elastoplastic «healing» of the impression after unloading is possible for a number of materials. Therefore, quantitative results obtained using this method of measurement are generally meaningless. No impression is left on a rubber sample after indentation, but this does not mean its infinitely high hardness.

Thus, all recent reports on materials much harder than diamond concern carbon materials with a nanocrystalline (sometimes partially amorphous) structure. The nanostructure indeed allows approaching hardness to the «ideal» value for elastic loading. In the case of cubic boron nitride, the formation of nanostructures increases hardness by a factor of about 2. Exact estimates for diamond are difficult. The formation of a nanostructure makes it possible to increase the measured hardness at least by 20–30% (from 80–100 to 120–130 GPa). At the same time, the absolute values of hardness of diamond materials of 130 GPa and above should be treated with caution. Even the most reliable results on Knoop hardness of 140 GPa reported by Irifune et al. were observed at a quite low load of 5 N, when the degree of elastic recovery was certainly high. Such values at the same loading parameters can be considered only as comparative when studying various materials. The samples synthesized by Blank's, Dubrovinsky's, and Tian's groups are very similar in structure to the samples obtained by Irifune et al. Nanopolycrystalline diamond samples with plate-like grains with a size of 10 nm are likely the hardest because the Vickers diamond indenter was destroyed at their indentation even at low loads [16]. Superhard carbon samples obtained by other groups were less hard because Vickers diamond indenters were not destroyed at their indentation. The nanotwinned structure of diamonds obtained by Chinese groups possibly promotes the highest degrees of elastic recovery of impression after unloading. Finally, we note that the «ideal» hardness of diamond even under fully elastic loading is 200–250 GPa [28]; consequently, any values of «recovered» hardness (after unloading) above 200 GPa are certainly meaningless. The absence of noticeable plastic deformation of the tip of Knoop, Vickers, or Berkovich diamond indenter after indentation also means that the hardness of the studied material did not exceed 180 GPa.

Thus, all announcements of carbon materials with hardness above 150 GPa should be considered as doubtful and the results of measurements of hardness in the range of 120–150 GPa should be treated with great caution as qualitative comparative data.

## 4. Are future quantitative *in situ* studies of the hardness of superhard materials possible?

Thus, reliable quantitative methods for the measurement of hardness above 100 GPa (100 GPa is the typical hardness of a diamond indenter) are absent. At the same time, there are unambiguous values of critical shear stresses for various loading types at which a crystal lattice becomes unstable. Such an «ideal» hardness in the regime of strongly nonlinear deformations can be calculated *ab initio* but still unfortunately with a low accuracy. After the loss of stability, a cascade of dislocations or coherent twin boundaries and a contact pressure appear simultaneously in the lattice and, correspondingly, the measured hardness decreases sharply [28]. If multiple barriers are formed for the motion and multiplication of dislocations, the hardness of the sample will approach the ideal shear strength for a given loading type. In particular, for a sapphire single crystal, the ideal hardness determined by elastic moduli for the Berkovich pyramid is 57 GPa, the ideal hardness in the nonlinear deformation region before the avalanche nucleation of dislocation loops ("pop-in" transition) is 47 GPa, and the measured hardness in the plastic deformation regime after the "pop-in" transition is 28 GPa [28]. The polishing of the sapphire surface increases both the concentration of surface dislocations and the measured hardness to 32 GPa. The formation of the nanocrystalline structure of sapphire can obviously approach the measured hardness in the plastic deformation regime to the «ideal» value up to 47 GPa [28]. For most materials, the *in situ* nanoindentation method allows the correct and unambiguous determination of both the «ideal» hardness (before the "pop-in" transition) and the real sample hardness after the "pop-in" transition, which is determined by the grain size and concentration of dislocations and other defects. For amorphous and polycrystalline materials the "pop-in" transition is not usually observed, but there are other methods to detect the plasticity onset. Thus, very recently the elastic-plastic transition has been studied for both single-crystal and polycrystalline cBN [84]. Materials with hardness close to the diamond value are exceptions because for them the onset of the elastoplastic transition usually cannot be observed and, correspondingly, it is impossible to determine the ideal shear strength of the material at a given loading type. Thus, the situation with the measurement of hardness above 80–100 GPa is unsatisfactory. Hardness can be studied only in terms of an impression after loading. In this case, impressions with a normal quality without plastic deformation and (or) fracture of the indenter are left only at moderate loads when the degree of elastic recovery of the impression is still high, and the measured hardness «is not saturated». Hardness values obtained in this case could be used to compare different materials, but this is not entirely correct. A material with lower elastic moduli and a lower ideal shear strength can simultaneously have a higher degree of elastic recovery (as rubber), and impressions left after indentation will formally correspond to a higher hardness. The comparison of the ideal shear strength of different superhard materials obviously requires *in situ* measurements.

We recall that the contact pressure at indentation by a spherical indenter increases with load and penetration depth, whereas the pressure at indentation by a conical or pyramidal ideally hard indenter is determined only by Young's modulus of material and the apex angle. In the case of real Vickers or Knoop indenter, the almost spherical part (rounding at the apex) penetrates into a material at the initial stage of indentation. Starting with a penetration depth of 0.2–0.4 μm, the penetration of the pyramidal part of indenter begins. The plastic deformation of materials close to diamond in hardness should begin at an indenter penetration depth higher than 0.4 μm (see nanoindentation data in [16]). In this case, plastic deformation regions in the sample are localized under the impression center (at a depth of ~0.1 μm under the contact) and near the pyramid edges (at a distance of about 20–50 nm from the contact places). Stresses corresponding to the plastic flow in a diamond-like material will obviously appear only at sufficiently high loads (~1 N), at which the deformation and fracture of the indenter begins.

To solve the problem of *in situ* study of the hardness of ultrahard materials, it is obviously necessary to return to the nanoindentation technique, but at a new technological level. First, the maximum best polishing of the studied material is required, possibly with fine finishing by an ion beam to a roughness of about 10–20 nm. Second, indenters should be made of defectless type IIa diamonds. The apex and edges of the indenter after polishing should also be «refined» to the maximum sharpness by ion beam. We note that diamond nanoneedles under loading demonstrate the highest (up to 9%) elastic strain without fracture [85]. The rounding radius at the apex and near the edges of indenter should be no more than 20–50 nm. The type and angle of the pyramid optimal for indentation should be selected at tests. Indentation should be performed in the slowest loading regime. In this case, the elastic-plastic transition in a ultrahard material should occur at a penetration depth of 100–200 nm at loads of ~100 mN. Such homemade indenters will possibly allow detecting the onset of plasticity *in situ* and experimentally determining both the «ideal» shear strength of ultrahard diamond and other carbon nanomaterials before the elastic-plastic transition and their hardness after the transition to the plastic regime.

## IV. Myth III: quantum nanoconfinement as a «theoretical» mechanism for the synthesis of superhard materials

The concept of the contribution of quantum confinement to the highest hardness of nanomaterials is perhaps the most irrational. It is caused by a stack of errors, incorrect treatments, and unjustified extrapolations of various theoretical and experimental results. It is known that a high concentration of defects and nanostructuring complicate processes of plastic deformation and improve the mechanical characteristics such as hardness, fracture toughness, and wear resistance (see [1, 2] and references therein). These effects are usually taken into account by means of empirical formulas (Hall–Petch law, etc.). However, in recent years, intensive efforts were focused on the development of the quantitative theoretical foundation for prediction of ways for improvement of mechanical characteristics of nanomaterials [32, 67]. In particular, an additional mechanism was proposed to increase the hardness of nanostructured materials owing to quantum confinement effects [67]. In this case, quantum confinement is a change in the electron wavefunctions and the corresponding change in the semiconductor gap and other physical properties of a material at the transition to the nanoscale.

The developed «theory» of the hardness increase owing to quantum confinement effects can be briefly formulated as follows

1. The activation energy of plastic deformation in semiconductors, including diamond-like ones, certainly correlates with the semiconductor gap [63, 64].

2. This energy is proportional to the hardness of a material (resistance to bond rupture); therefore, hardness is related to the gap width [29].

3. For nanoparticles of many semiconductor, the semiconductor gap or exciton energy increases with decreasing particle sizes [86-92] owing to quantum confinement effects (restriction of the scale of the wavefunction).

4. The comparison of the Kubo–Galperin theory [93] for nanoparticles with experimental data for a number of semiconductors makes it possible to reveal a universal relation between an increase in the semiconductor gap and nanocrystal size [67].

5. This relation, along with the contribution from the Hall–Petch effect, was further used to develop the «microscopic theory of hardness» for nanomaterials [32].

6. This relation was also used to explain an anomalously high hardness of nanotwinned cubic boron nitride and nanotwinned diamond [76, 77].

7. It was predicted that the contribution from confinement effects to hardness for the smallest twins in diamond is about 300 GPa; this contribution, together with the possible contribution from the Hall–Petch effect (about 200 GPa), should result in the possible hardness of nanotwinned diamonds up to 600 GPa [77–81].

At first glance, all these items seem convincing, but in reality, each item contains errors and discrepancies, and some of them are totally unacceptable.

At the beginning of the 1970s, it was noticed that the activation energy of plastic deformation in covalent crystals with a diamond-like lattice is indeed close to the doubled semiconductor gap (see [63] and references therein). In fact, this means that the nucleation of a dislocation loop requires the rupture of a covalent bond, which in turn means the excitation of a valence electron to the conduction band. Later, this simplified picture was constructively criticized [65, 66]. It was found that a simple relation between the activation energy of plastic flow and the gap width is invalid for narrow-gap semiconductors: the activation energy remains finite when the gap vanishes [65]. Then, it was established that Pugh's ratio $G/B$ should also be taken into account to correctly describe the activation parameters of plasticity [66]. Most works in this field focused on the temperature dependence of hardness at high temperatures rather than on its absolute value, although correlation between hardness at moderate temperatures and the semiconductor gap was also mentioned. Thirty years later, Gao et al. [29] returned to the idea of correlation between hardness and the gap for semiconductor crystals. In this approach, the process of indentation serves as a high temperature: the rupture of bonds occurs in the plastic deformation regime and hardness is approximately proportional to the energy of covalent bonds. This model is certainly applicable only to a very narrow class of materials and does not describe a high hardness of metallic borides and carbides. This correlation (for a narrow class of materials) is generally justified and is not worse than others. Indeed, a high degree of covalence and a large gap mean a high electron density (and, therefore, high elastic moduli) and a high degree of localization of the electron density (and, therefore, high Pugh's ratio $G/B$) [2]. Further, ideas of Gao et al. on the relation of hardness to the semiconductor gap and ideas of Chen [30] on the importance of Pugh's ratio $G/B$ were combined and developed in a number of works (see, e.g. [31]). The only problem is that all these correlations are not strict, and formulas based on them should be treated as empirical.

At the same time, the recognition of the relation of mechanical characteristics of covalent crystals to the semiconductor gap stimulated new ideas on the mechanism of the nanostructure effect on the materials hardness. The paper [67] is remarkable in this respect. To explain the increase in the hardness of nanocrystalline diamonds as compared to ordinary single crystals, the authors considered the possibility of increasing the semiconductor gap owing to quantum confinement effects instead of an ordinary mechanism of the dislocations drag such as Hall–Petch effect.

To calculate an increase in the gap in an individual nanocrystal, the "Kubo theory" is used [93]. This is at least strange. First, the authors of [93] considered the electronic levels of metallic nanoparticles, and the formula presented in [67] is the energy spacing between the levels, which is equal to the Fermi energy divided by the number of atoms. This quantity is very small (~1 meV) even for 1-nm particles, and the effects considered in [93] are nothing to do with of quantum confinement in semiconductors and can be manifested in metallic nanoparticles only at very low temperatures. Second, this energy gap is inversely proportional to the nanoparticle volume ($\delta \sim 1/V$) [67, 93]. Third, at the transition from Eq. (4) to Eq. (5) in the paper [67], an unexplained change from $\delta$ to $\delta_P$ has been made. These two values differ only in the «artificially introduced» dimensionless parameter $K$, which moreover indefinitely depends on the nanoparticle size. Then, the authors of [67] made approximation using a number of heterogeneous data [89-91] on the dependence of excitons energy and semiconductor gap on the nanocrystals size (Fig. 3 in [67]). As a result, the authors of [67] concluded that an increase in the semiconductor gap energy for semiconductor nanocrystals can be described by a certain universal function and this increase is approximately proportional to the size of nanocluster rather than to its volume. We note that most of the data shown in Fig. 3 in [67] were taken from [90], where an increase in the exciton energy in CdS clusters at decreasing cluster size was observed, which is indirectly due to an increase in the semiconductor gap. However, the main problem is different. In reality, this theoretical consideration in [67] has no connection to the theory of quantum confinement in semiconductor materials.

The concept of quantum confinement was known since the middle of the past century. The general idea of confinement is that the energy spectrum of carriers can change because of quantum confinement when the spatial scale of the electron wavefunction (or the de Broglie wavelength of the carrier) becomes comparable to a certain size of the particle (cluster). Quantum confinement effects in semiconductors become significant when the Bohr radius of an exciton (bound state of an electron and a hole) becomes comparable with the size of nanoparticles; i.e. the effect is to a certain degree threshold but smooth. At a large dielectric constant and a small effective mass of electrons, this corresponds to a particle size of several nanometers. An accurate theoretical description of the energy shift caused by quantum confinement effects is rather difficult [94–97]. In any case, the increase in the effective semiconductor gap in a nanocluster is not attributed to the Kubo theory and is not described at all by equations similar to Eq. (4) presented in [67]. The leading contribution to the energy shift obviously comes from the term corresponding to the kinetic energy, which is inversely proportional to the square rather than to the cube of the cluster size (as in Eq. (4) in [67]) or to the cluster size (as in Eq. (6) in [67]). The correct kinetic quantum contribution differs from Eq. (4) in [67] in a factor of $R/a$, where $R$ is the nanoparticle size and $a$ is the interatomic distance.

The concept of quantum confinement was rapidly developed at the beginning of the 1990s in view of the synthesis and study of porous silicon [86-88]. For porous silicon samples, which are aggregates of weakly coupled filament-like nanoclusters, strong luminescence was observed, and the valence and conduction bands were noticeably shifted relative to the bulk silicon samples. One of the explanations of the observed effects was reduced to the quantum confinement of electrons (and holes) in porous silicon [86-88]. Subsequent studies indicated that silicon nanoclusters obtained by other methods also demonstrate an increase in the exciton energy and in the effective semiconductor gap with decreasing cluster size [80]. An increase in the gap for ultrasmall clusters (~1 nm) is about 1 eV. It is noteworthy that data of different works are contradictory in view of the complexity of characterization and study of nanoparticles, as well as of chemical impurities inevitably presenting on the surface of clusters. One of the most detailed studies of the influence of quantum confinement effects on the exciton energy was performed for CdS semiconductor clusters with sizes from 1 to 6 nm [90]. It was found that the semiconductor gap for the smallest clusters increases by about 1 eV, i.e., about 30% compared to the bulk semiconductor. Theoretical description of the results obtained for CdS was proposed in [90] and in subsequent work [97]. Nanoconfinement effects for diamonds were analyzed only in a few works for individual clusters [91] and island-like nanofilms [92]. A noticeable increase in the gap (by about 10–20%) was observed only for ultrasmall (1–2 nm) clusters [91, 92]. In this case, a large contribution obviously comes from surface atoms whose number is comparable to the number of atoms inside nanoparticles. In the case of diamond nanoparticles, the structure on the surface of clusters is transformed to a fullerene-like structure, which also affects the observed effects [91].

For all studied semiconductor nanocrystals, a noticeable increase in the gap is observed starting with 1–3-nm clusters, and the maximum increase in the gap is about 1 eV. The gap increase in this case is not directly connected with the complexity of plastic deformation. Dislocations in nanocrystals with such sizes can hardly be formed [98], and a nanoparticle cluster demonstrates a super-elastic behavior even at very high pressures [99].

Moreover, quantum confinement effects are observed only for individual nanoparticles or for weakly coupled clusters. Boundary conditions for the electron wavefunction at the grain boundary in a nanopolycrystal are significantly different from those for the free surface, and quantum confinement effects were not observed in any bulk semiconductor nanopolycrystal. In this respect, coherent twin boundaries considered in [76–81] are specific because structural disorder effects at such twin boundaries are minimal, and no quantum confinement effects can occur. Furthermore, twin boundaries themselves can be involved in plastic deformation under loading.

As mentioned above, empirical formulas for quantum confinement effects were added in subsequent works to the «theory» of hardness [32], which is said to explain «observed» hardness of 200–300 GPa and to predict the possibility of a further increase in the hardness of nanomaterials up to 600 GPa [77–81].

The absence of any noticeable quantum confinement effects in bulk nanopolycrystals, particularly with coherent intergrain boundaries, is certainly a serious reason against new theories of hardness. However, this reason is not necessary in this case. The matter is even more curious. «Fundamental» empirical formulas (6) and (7) in [67] for an increase in the gap and for a corresponding increase in hardness include the cluster size $D$ in angstroms. The same expressions were exactly rewritten in the basic work of the Chinese group [32] (Eqs. (18) and (19)) and were used in all subsequent works on superhard materials, but the nanocrystallite size $D$ in all these works was erroneously presented in nanometers rather than in angstroms. This means that even the hypothetical contribution from confinement was overestimated by an order of magnitude. According to expressions obtained in [67], the maximum hypothetical increase in the gap and hardness in the synthesized nanotwinned diamond at a twin thickness of 5 nm (= 50 Å) [77] cannot exceed 5%. As mentioned above, an increase in the gap even for ultrasmall diamond clusters with a size of ~1 nm is no more than 20% [91, 92]. We note that the contribution from the quantum confinement effect to hardness in the original work [67], where the quantum confinement mechanism of increasing hardness was firstly proposed, was indeed estimated as 10–20% at minimum crystal sizes of 1–1.5 nm.

As a result, the bright model of the effect of quantum confinement on the hardness of semiconductor materials presented at the beginning of this section must be corrected as follows

1. The activation energy of plastic deformation for a number of covalent crystals correlates with the semiconductor gap, but this correlation is not strict.

2. Correlation between the measured hardness and semiconductor gap is observed for certain classes of semiconductor materials, but this correlation is even less strict than that mentioned in item 1).

3. In many semiconductor nanoparticles (clusters), the exciton energy and gap width increase with a decrease in the cluster sizes to nanometers. This increase is observed only for individual or weakly coupled clusters and is never observed for a bulk nanocrystalline material.

4. The increase in the gap has a complex dependence on the cluster size, as well as on the interaction with surface atoms, and is not described by the Kubo–Galperin theory.

5. There are no reasons for the use of data on quantum nanoconfinement in small clusters to explain a possible increase in hardness in nanopolycrystals.

6. The hypothetical effect of quantum confinement on the hardness of nanotwinned diamond and cubic boron nitride is overestimated by an order of magnitude because of the random (or systematic) change in units of the cluster size measurement.

7. Nanostructuring indeed allows increasing hardness by a factor of 1.5–2 for most materials and by 20–40% for diamond, but only owing to the long known Hall–Petch effect.

Thus, there are no reliable scientific reasons for an additional hardness increase in nanopolycrystals caused by the quantum confinement effects.

## V.  Conclusions

Summarizing the above discussions leads to the conclusion that under normal conditions diamond remains the hardest known material. The fundamental impossibility of synthesizing materials with elastic moduli and mechanical properties noticeably exceeding those of diamond has been demonstrated. Accordingly, any statements on the possibility of producing materials with elastic characteristics and/or hardness several times higher than the corresponding values for diamond cannot be considered as scientifically reliable.

**Acknowledgements**

The authors thank Prof. T. Irifune, Dr. A.G. Lyapin, Prof. S.M. Stishov, Dr. S.N. Dub, Dr. O.B. Tsiok and Prof. V.N. Ryzhov for useful discussions. V.V.B. is grateful to the Russian Science Foundation for financial support.

ORCID IDs : Vladimir L. Solozhenko  https://orcid.org/0000-0002-0881-9761